\documentclass[twocolumn,showpacs,pra,12]{revtex4}% Physical Review A
\usepackage{amsfonts, amssymb,amsmath} % Include mathematical symbols
\usepackage{graphicx}% Include figure files
\usepackage{dcolumn}% Align table columns on decimal point
\usepackage{bm}% bold math
\newcommand{\be}{\begin{equation}}
\newcommand{\ee}{\end{equation}}
\newcommand{\bea}{\begin{eqnarray}}
\newcommand{\eea}{\end{eqnarray}}

\newcommand{\ti}{\i}
\newcommand{\tu}{\"u}
\newcommand{\tc}{\c c}

\newcommand{\tg}{\u g}

\newcommand{\tro}{\"o}

\newcommand{\tU}{\"U}
\newcommand{\tI}{\.I}
\newcommand{\Sch}{Schr\tro dinger equation~}

\newcommand{\ddf}{Dirac delta functions~}

\newcommand{\cf}{condensate fraction~}

\newcommand{\del}{$\delta$~}

\begin{document}

%\preprint{APS/123-QED}

\title{Bose-Einstein condensate in a harmonic trap decorated with Dirac \del functions}

\author{Haydar Uncu$^1$, Devrim Tarhan$^2$, Ersan Demiralp$^{1,3}$,
\"{O}zg\"{u}r E. M\"{u}stecapl{\i}o\~{g}lu$^4$}
\affiliation{$^{1}$Department of Physics, Bo\tg azi\tc
 i University, Bebek, 34342, \tI stanbul, Turkey}
\affiliation{$^2$Department of Physics, Harran University,
Osmanbey Yerle\c{s}kesi , \c{S}anl\i urfa, Turkey}
\affiliation{$^{3}$Bo\tg azi\tc i University-T\tU B\tI TAK Feza
G\tu rsey Institute
 Kandilli, 81220, \tI stanbul, Turkey}
 \affiliation{$^4$
Department of Physics, Ko\c{c} University Rumelifeneri yolu,
Sar\ti yer, 34450, \tI stanbul, Turkey}
\date{\today}
%-------------------------------------------------------------------------------
\begin{abstract}
We study Bose-Einstein condensation in a harmonic trap with a
dimple potential. We specifically consider the case of a tight and
deep dimple potential which is modelled by a Dirac \del function.
This allows for simpler, explicit numerical and analytical
investigations of noninteracting gases. Thus, the \Sch is used
instead of the Gross-Pitaevski equation. Calculating the atomic
density, chemical potential, critical temperature and condensate
fraction, the role of the relative depth of the dimple potential
with respect to the harmonic trap in large condensate formation at
enhanced temperatures is clearly revealed.
\end{abstract}
%---------------------------------------------------------------------
%------------------------------------------------------------------------
\maketitle
%\narrowtext
%==============================================================================

\section{INTRODUCTION}

Bose-Einstein condensation was the last major discovery of
Einstein \cite{einstein,bose}. By applying the Bose-Einstein
statistics to an ideal gas, Einstein showed that ``from a certain
temperature on, the molecules condense without attractive forces
... " \cite{pais} and discovered the Bose-Einstein condensation in
1925. This theoretical prediction motivated the experimental
studies for realizations of Bose-Einstein condensates of gases.
Seventy years after the prediction of Einstein, Bose-Einstein
condensates (BEC) of dilute gases have been observed at very low
temperatures  by using ingenious experimental designs
\cite{anderson,bradley,davis}.

Experimentally available condensates are systems with finite
number of atoms $N$, confined in spatially inhomogeneous trapping
potentials. Their understanding requires theories that go beyond
the usual treatments based upon London's continuous spectrum
approximation \cite{london} or the thermodynamic limit
$N\rightarrow \infty$. Such studies \cite{ketterle,druten} reveal
that BEC can occur in harmonically trapped lower-dimensional
systems for finite $N$ despite the enhanced importance of phase
fluctuations \cite{khawaja}. Quasicondensates with large phase
fluctuations may still occur \cite{petrov}. This is in contrast to
standard results \cite{groot}, in agreement with the
Mermin-Wagner-Hohenberg theorem \cite{mermin,hohenberg} in the
thermodynamic limit. Bose-Einstein condensation in one dimension
(1D) with a harmonic trap is attractive due to enhanced critical
temperature and condensate fraction \cite{ketterle,druten}.
Recently, one- and two- dimensional BECs have been created in
experiments\cite{gorlitz}. One-dimensional condensates were also
generated on a microchip \cite{ott,hansel} and in lithium mixtures
\cite{schreck}.

Modification of the shape of the trapping potential can be used to
increase the phase space density \cite{pinkse}. ``Dimple"-type
potentials are the most favorable potentials for this purpose
\cite{kurn,weber,ma}. The phase-space density can be enhanced by
an arbitrary factor by using a small dimple potential at the
equilibrium point of the harmonic trapping potential \cite{kurn}.
A recent demonstration of caesium BEC exploits a tight dimple
potential \cite{weber}. Quite recently, such potentials were
proposed for efficient loading and fast evaporative cooling to
produce large BECs \cite{comparat}. Tight dimple potentials for
one-dimensional (or strictly speaking quasi-one-dimensional) BECs
offer attractive applications, such as controlling interaction
between dark solitons and sound \cite{parker}, introducing defects
such as atomic quantum dots in optical lattices \cite{jaksch}, or
quantum tweezers for atoms \cite{diener}. Such systems can also be
used for spatially selective loading of optical lattices
\cite{griffin}. In combination with the condensates on atom chips,
tight and deep dimple potentials can lead to rich novel dynamics
for potential applications in atom lasers, atom interferometers
and in quantum computations (see Ref. \cite{proukakis} and
references therein).

In this paper, we investigate Bose-Einstein condensation in a 1D
harmonic trap decorated by a tight and deep dimple potential. The
quantum kinetics of a similar system but for a 3D single-mode
trap, modelled by a deep but narrow spherical square potential
well, was studied in Ref. \cite{anglin}, where a master equation,
to describe condensate growth, was developed.  We model the dimple
potential using a Dirac \del function. The delta function can be
defined via a Gaussian function \cite{lighthill}
$g(x,a)=(1/\sqrt{\pi} a)\exp{(-x^2/a^2)}$ of infinitely narrow
width $a$ so that $g(x,a)\rightarrow\delta(x)$ for $a\rightarrow
0$. This allows for analytical calculations in some limiting cases
as well as a simpler numerical treatment for arbitrary parameters.
We calculate the transition temperature as well as the chemical
potential and condensate fraction for various number of atoms and
for various relative depths of the dimple potential. For
describing a system with interacting particles, the
Gross-Pitaevski equation is usually utilized. We note that we
neglect the interactions between atoms in our model, and thus the
\Sch for the harmonic trap with the dimple potential is solved.

The paper is organized as follows. In Sect. II,  we present the
analytical solutions of the \Sch for a Dirac $\delta$-decorated
harmonic potential and the corresponding eigenvalue equation. In
Sect. III, determining the eigenvalues numerically, we show the
effect of the dimple potential on the condensate fraction and the
transition temperature. In Sect. IV, we present a semiclassical
method to calculate condensate fraction when a dimple potential is
added to harmonic trap adiabatically.  Analytical results in the
limit of strong dimple potential are presented in the Sect. V.
Finally, we present our conclusions in Sect. VI.

%-------------------------------------------------------------------------------

\section{Harmonic potential Decorated with \ddf}

We begin our discussion with the one dimensional harmonic
potential decorated with the Dirac \del functions
\cite{demiralp2}-\cite{demiralp3}. This potential is given as
\be V(x)=\frac{1}{2} m \omega^2 x^2-\frac{\hbar^2}{2m} \sum_i^P
\sigma_i \delta(x-x_i), \label{potential} \ee
where $\omega$ is the frequency of the harmonic trap, $P$ is a
finite integer, and $\sigma_i$'s are the strengths (depths) of the
dimple potentials located at $x_i$'s with $x_1<x_2<...<x_P$ with
$x_i \in (- \infty,\infty)$. The factor $\hbar^2/2m$ is used for
calculational convenience. A negative $\sigma_i$ value represents
repulsive interaction while positive $\sigma_i$ value represents
attractive interaction. We can write  time-independent \Sch as
\be -\frac{\hbar^2}{2m} \frac{d^2 \Psi(x)}{dx^2} + V(x) \Psi(x)= E
\Psi(x). \label{Scheq} \ee
By inserting $E=(\xi+\frac{1}{2}) \hbar \omega$, with $\xi$ a real
number, and introducing dimensionless quantities $z=x/x_0$, and
$z_i=x_i/x_0$ with $x_0=\sqrt{\hbar/2m \omega}$, the natural
length scale of the harmonic trap, we can re-express Eq.
(\ref{Scheq}) as
\be
 \frac{d^2 \Psi(z)}{dz^2}+ \left [ \xi + \frac{1}{2}-\frac{z^2}{4}
 +  \sum_i^P \Lambda_i \delta(z-z_i)\right ] \Psi(z) =0,
 \label{parcyldiffeq}
\ee
where $\Lambda_i=x_0\sigma_i$. For $z\neq z_i$, the Eq.
(\ref{Scheq}) has two linearly independent solutions. For $\xi
\neq 0,1,2,... $ these linearly independent solutions are
parabolic cylinder functions $D_{\xi}(z)$ and $ D_{\xi}(-z) $,
defined as:
 \bea
 & & D_{\xi} (z)  = 2^{\frac{\xi}{2}} e^{-\frac{z^2}{4}}
 \times
 \nonumber
\\ & &
\left \{ \frac{\sqrt{\pi} \Phi \left( -\frac{\xi}{2},\frac{1}{2};
\frac{z^2}{2} \right)}{\Gamma\left(\frac{1- \xi}{2} \right)}  -
\frac{\sqrt{2 \pi} z \Phi \left( \frac{1-\xi}{2},\frac{3}{2};
\frac{z^2}{2} \right)}{\Gamma\left(-\frac{ \xi}{2} \right)}
\right\} \label{Posparcyl}
 \eea
\bea & & D_{\xi}(-z)= 2^{\frac{\xi}{2}} e^{-\frac{z^2}{4}} \times
\nonumber
\\ & &
 \left \{ \frac{\sqrt{\pi} \Phi \left(
-\frac{\xi}{2},\frac{1}{2}; \frac{z^2}{2}
\right)}{\Gamma\left(\frac{1- \xi}{2} \right)}+ \frac{\sqrt{2\pi}
z \Phi \left( \frac{1-\xi}{2},\frac{3}{2}; \frac{z^2}{2}
\right)}{\Gamma\left(-\frac{\xi}{2} \right)} \right\}  ,
\label{Negparcyl}
 \eea
where $\Phi \left( -\xi / 2, 1/2; z^2/2 \right)$ and $\Phi \left(
(1-\xi)/2, 3/2; z^2/2 \right)$ are the confluent hypergeometric
functions $\Phi \left( a,b; z^2/2 \right)$ defined in Ref.
\cite{lebedev}
 for $a= \xi/2$,  $b=1/2$  and $a=(1-\xi)/2$, $b=3/2$, respectively.
Thus, the general solution of Eq. (\ref{parcyldiffeq}) between the
locations of two Dirac \del functions is
\be \Psi(z)= a_i \, D_{\xi}(z) +  b_i D_{\xi}(-z), \label{gensol}
\ee
with $a_i$ and $b_i$ can be written in terms of $b_1$ via transfer
matrix method and $b_1$ is determined by normalization
($a_1=b_{P+1}=0$ due to square integrability) \cite{demiralp3,
demiralp}. $D_{\xi}(-z)$ is regular as $z\rightarrow - \infty$,
but $|D_{\xi}(-z)| \rightarrow + \infty$ as $z \rightarrow \infty
$, and $D_{\xi}(z)$ is regular as $z\rightarrow \infty$, but
$|D_{\xi}(z)| \rightarrow + \infty $ as $z \rightarrow - \infty$
\cite{lebedev}. By using the transfer matrix method explained in
Refs. \cite{demiralp2,demiralp}, one finds the eigenvalue equation
for the potential given in Eq. (\ref{potential}). For the special
case of the potential given in Eq. (\ref{potential}), harmonic
potential decorated with one Dirac \del function at
its center $P=1, \; x_1=0 $,    %
\be V(x)=\frac{1}{2} m \omega^2 x^2-\frac{\hbar^2}{2m}  \sigma
\delta(x), \label{onedelta}\ee
the eigenvalue equation of the even states is written as
\cite{avakian,demiralp3,mistake}:
\be \frac{\Gamma((1-\xi)/2)}{\Gamma(-\xi/2)}= \frac{\Gamma(3/4-E/2
\hbar \omega)}{\Gamma(1/4-E/2 \hbar \omega)} = \frac{\sigma
\sqrt{\hbar/m \omega}}{4}. \label{eigenvaleq} \ee
In this case, the odd solutions of the harmonic potential are
unaffected because the Dirac \del function is located at the
center of the harmonic potential. On the other hand, the energy
eigenvalues of even states change as a function of $\sigma$. The
ground state energy eigenvalue decreases unlimitedly as $\sigma$
increases (attractive case). However, the energies of the excited
even states are limited by the energies $E_{2n+1}$ of odd states
and as $\sigma \rightarrow \infty$, $E_{2n+2} \rightarrow
E_{2n+1}=(2n+1+1/2) \hbar \omega$ where  $n=0,1,...$. Thus, as
$\sigma \rightarrow \infty$ the energy eigenvalues of these states
go to the energy eigenvalue of the one lower eigenstate, and the
odd energy eigenstates asymptotically become doubly degenerate.

%%%%%%%%%%%%%%%%%%%%%%%%%%%%%%%%%%%%%%%%%%%%%%%%%%%%%%%%%%%%%%%%%%%%%%%%%%%%%%%
%%%%%%%%%%%%%%%%%%%%%%%%%%%%%%%%%%%%%%%%%%%%%%%%%%%%%%%%%%

\section{BEC In a One-Dimensional Harmonic Potential with a Dirac \del Function}

We begin our discussion about BEC in a one-dimensional harmonic
potential decorated with a Dirac \del function by investigating
the change of the critical temperature as a function of $\sigma$.
One estimates a  $\sigma$ value using the parameters of
Ref.\cite{ma}. In this paper, the minimum value of the dimple
potential is given as $ U_c=k_B \, 4 \, \mathrm{\mu K} $ and
average potential width is given as $r=1-100 \,  \mathrm{\mu m} $.
We equate the strength ($-\hbar^2 \sigma /2m $) to the product
$U_c \, r$ to get an estimate of $\sigma$. We find that $\sigma$
varies approximately between $ 10^8 \, \mathrm{1/m} $ and $10^{10}
\, \mathrm{1/m} $ as r changes from $ 1 \, \mathrm{\mu m} \,
\mathrm{to} \, 100 \, \mathrm{\mu m}$. We define a dimensionless
parameter in terms of $\sigma$ as:
\be \Lambda= \sigma \sqrt{ \frac{\hbar}{m \omega}}\, .
\label{lambda} \ee
If $ 10^8 \, 1/ \mathrm{m} \leq \sigma \leq 10^{10} \, 1/
\mathrm{m} $ then $ 460 \leq \Lambda \leq 46000$ for the
experimental parameters m$=23$ amu ($^{23}\mathrm{Na}$), $\omega=
2 \, \pi  \times \, 21$ Hz \cite{hau} and $ 230 \leq \Lambda \leq
23000$ for the experimental parameters m$=133$ amu
($^{133}\mathrm{Cs}$), $\omega= 2 \, \pi \, \times 14$ Hz
\cite{weber}. In this work, we  show that, even for small
$\Lambda$ values, the \cf and critical temperature change
considerably.

\begin{figure}
{\vspace{0.5cm}}
\includegraphics[width=3.5 in]{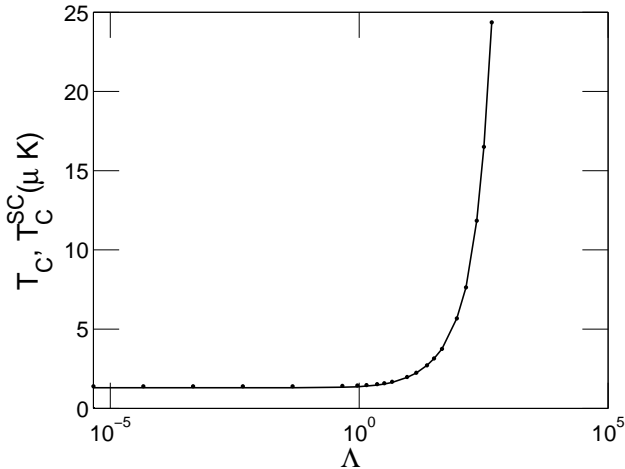}
\caption{ The critical temperature $T_c$ and $T_c^{SC}$   vs
$\Lambda $ for $N=10^4$. $\Lambda $ is a dimensionless variable
defined in Eq. (\ref{lambda}). The crosses and dots show $T_c$ and
$T_c^{SC}$ values, respectively. Here we use m $=23$ amu
($^{23}\mathrm{Na}$) and $\omega= 2 \pi \times 21$ Hz \cite{hau}.
The logarithmic scale is used for $\Lambda$ axis.} \label{critemp}
\end{figure}

The critical temperature ($T_c$) is obtained by taking the
chemical potential equal to the ground-state energy
($\mu=E_g=E_0$) and
\be N \approx \sum_{i=1}^{\infty} \frac{1}{e^{\beta_c
\varepsilon_i}-1} \;\; , \label{Tcritical} \ee
at $T=T_c$, where $\beta_c=1/(k_B T_c)$. For finite $N$ value, we
define $T_c^0$ as the solution of Eq. (\ref{Tcritical}) for
$\Lambda=0$ (only the harmonic trap).

In Eq. (\ref{Tcritical}),  $ \varepsilon_i$'s are the eigenvalues
for the potential given in Eq. (\ref{onedelta}). The energies of
odd states are unchanged and equal to $(2n+1+1/2) \hbar \omega$.
The energies of even states are found by solving Eq.
(\ref{eigenvaleq}) numerically. Then, these values are substituted
into the Eq. (\ref{Tcritical}), and finally this equation is
solved numerically to find $T_c$. We obtain $T_c$ for different $
\Lambda $ and show our results in Fig. \ref{critemp}. In this
figure, logarithmic scale is used for $\Lambda$ axis. As $\Lambda$
increases, the critical temperature increases very rapidly when $
\Lambda > 1$. The solid line in Fig. \ref{critemp} shows the
change of the critical temperature with $\Lambda$. Here we take
$N=10^4$ and use the experimental parameters m$=23$ amu
($^{23}\mathrm{Na}$) and $\omega= 2\pi \times 21$ Hz \cite{hau}.

\begin{figure}
\centering{\vspace{0.5cm}}
\includegraphics[width=3.5 in]{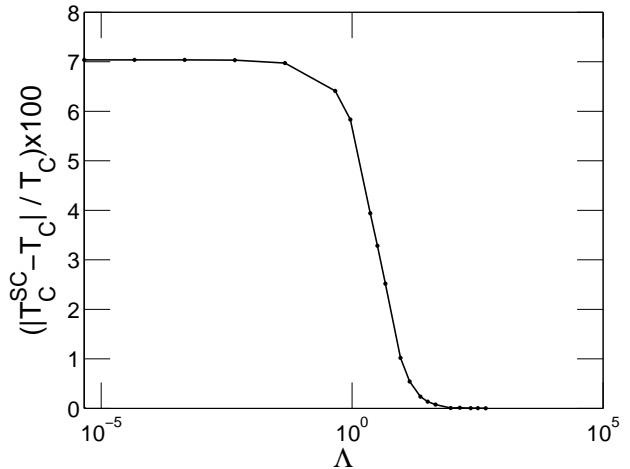}
\caption{ ($T_c^{SC}-T_c/T_c) \times 100$ vs $\Lambda$, the
difference of the critical temperatures calculated by the
semiclassical and our method with respect to $\Lambda$. The
logarithmic scale is used for $\Lambda$ axis.} \label{error}
\end{figure}

We have calculated the change in the critical temperature
approximating dimple-type potentials  by Dirac \del function. The
change in the critical temperature can also be calculated  by
assuming that the optical trap is added to harmonic trap
adiabatically \cite{kurn,stringari}. In this semi-classical
approach, one uses the density of states and assumes the energy
spectrum is continuous utilizing the fact that $ \hbar \omega
/(k_B T) \ll 1$. For a one dimensional harmonic trap, the number
of particles in the thermal gas can be found as
\be
N= N_{0}+\frac{1}{\hbar \omega} \int_{(3/2) \, \hbar
\omega}^{\infty} d\varepsilon \frac{1}{e^{\beta (\varepsilon-
\mu)}-1} \, . \label{thermal gas}
\ee
where $ N_0$ denotes the average number of particles in the ground
state. The $ 1/(\hbar \omega) $ factor before the integral in Eq.
(\ref{thermal gas}) is the density of states for one dimensional
harmonic potential. The critical temperature $T_c^0$ in the
harmonic trap can be found by taking $\mu=\hbar \omega/2$ and
writing the following equation for the total number of particles,
N:
\be
N= \frac{1}{\hbar \omega} \int_{(3/2) \, \hbar \omega}^{\infty}
d\varepsilon \frac{1}{e^{\beta_c^0 \left(\varepsilon- \frac{1}{2}
\hbar \omega\right)}-1} , \label{critemphar}
\ee
where $\beta_c^0=1/(k_B T_c^0)$. The addition of an attractive
dimple potential to the harmonic trap decreases the ground-state
energy, resulting a decrease in the chemical potential. Since the
chemical potential is assumed to be equal to the ground sate
energy at critical temperature, replacing $\mu$ in Eq.
(\ref{thermal gas}) by $E_g$, the ground state energy for the
harmonic trap together with the dimple potential, it is possible
to calculate the critical temperature of the new system;
\be
N= \frac{1}{\hbar \omega} \int_{(3/2) \, \hbar \omega}^{\infty}
d\varepsilon \frac{1}{e^{\beta_c^{SC} (\varepsilon- E_g)}-1} .
\label{critempop}
\ee
where $\beta_c^{SC}=1 /(k_B T_c^{SC}) $ denotes the critical
temperature values obtained by semiclassical method.  Here we
assume that the energies of the excited state are same with the
harmonic trap by following Ref. \cite{kurn}. We have calculated
the critical temperature for various dimple potentials differing
according to potential depths. We take the $E_g$ values as the
ground state of the potential given in Eq. (\ref{onedelta}) for
different $\sigma$ [ or $\Lambda= \sigma \sqrt{ \hbar  / (m
\omega)}$ ] values  in order to compare the results of this method
with our results.

\begin{figure}
\centering{\vspace{0.5cm}}
\includegraphics[width=3.5in]{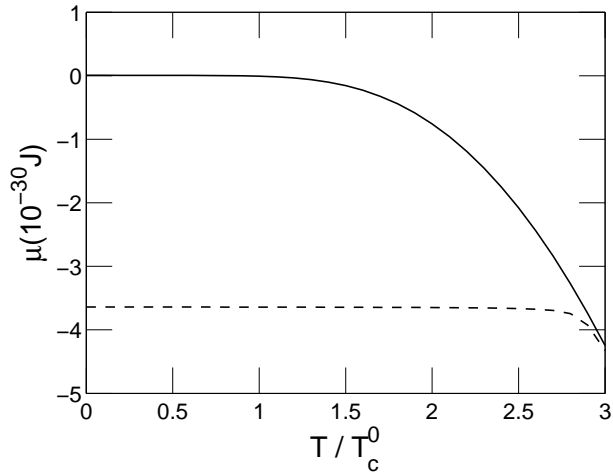}
\caption{The chemical potential $\mu$ vs temperature $T/T_c^0$ for
$N=10^4$ and solid line for $\Lambda=0$ and dashed line for
$\Lambda=46$. The other parameters are the same as Fig.
\ref{critemp}.}
 \label{muvstgraph}
\end{figure}

The critical temperature values calculated using this
semiclassical approach ($T_c^{SC}$) is shown in Fig. \ref{critemp}
as points. As seen from the figure, the critical temperature
values estimated by our method and by the semiclassical method are
almost equal. We show in Fig. (\ref{error}) percentage difference
of the critical temperatures that are calculated by using these
methods. The difference of these two methods for very small
$\Lambda$ values is due to the accurate calculation of the
summation with accurate energy values for the excited states for
our method; however, the summation is approximated with an
integral by using the harmonic trap density of states for the
semiclassical method.

\begin{figure}
\centering{\vspace{0.5cm}}
\includegraphics[width=3.5in]{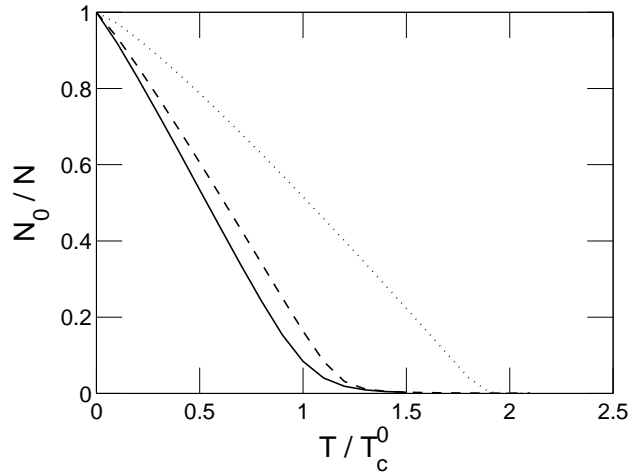}
\caption{$N_0/N$ vs $T/T_c^0$ for $N=10^6$ and solid line for
$\Lambda=0$, dashed line for $\Lambda=4.6$, dotted line for
$\Lambda=46$. The other parameters are the same as in Fig.
\ref{critemp}.} \label{condfrac}
\end{figure}

For a gas of N identical bosons, the chemical potential $\mu$ is
obtained by solving
\be N=\sum_{i=0}^{\infty} \frac{1}{e^{\beta (\varepsilon_i -
\mu)}-1}= N_0 + \sum_{i=1}^{\infty} \frac{1}{e^{\beta
(\varepsilon_i - \mu)}-1}, \label{chem pot} \ee
at constant temperature and for given N, where $\varepsilon_i$ is
the energy of state $i$. We present the change of $\mu $ as a
function of $T/T_c^0$ in Fig.\ref{muvstgraph} for $ N=10^4 $; $
\Lambda = 0$ and $ \Lambda = 46 $. By inserting $\mu $ values into
the equation
\be N_0= \frac{1}{e^{\beta (\varepsilon_0 - \mu)}-1},\ee
we find the average number of particle in the ground state.
$N_0/N$ versus $T/T_c^0$ for $N=10^6$ and $\Lambda= 0,  4.6 , 46 $
are shown in Fig. \ref{condfrac}. In this figure, the result for
$\Lambda=0$ is the same as the result obtained by Ketterle and van
Druten \cite{ketterle}. As mentioned in Ref. \cite{ketterle}, the
phase transitions due to discontinuity in an observable macro
parameter occurs only in thermodynamic limit, where $N \rightarrow
\infty$. However, we make our calculations for a realistic system
with a finite number of particles. Thus, $N_0/N$ is a finite
nonzero quantity for $T<T_c$ without having any discontinuity at
$T=T_c$.

\begin{figure}
\centering{\vspace{0.5cm}}
\includegraphics[width=3.5in]{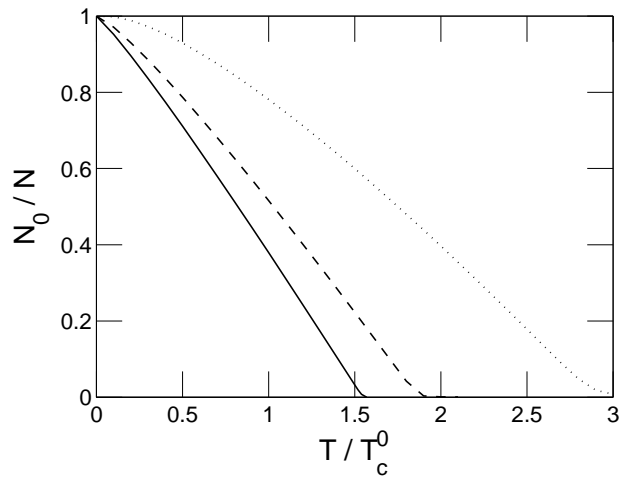}
\caption{Condensate $N_0/N$ vs temperature  $T/T_c^0$ for
$\Lambda=46$. The solid line for $N=10^8$, dashed line for
$N=10^6$, dotted line for $N=10^4$. The other parameters are the
same as in Fig. \ref{critemp}.} \label{condfrac2}
\end{figure}

It is useful to know the behavior of the condensate fraction as a
function of temperature for a fixed value of $\Lambda$. We present
the condensate fraction for $N=10^4,10^6,10^8$ when $\Lambda =46$
in Fig. \ref{condfrac2}. All condensate fractions for different N
values are drawn by using their corresponding $T_c^0$ values.
These $T_c^0$ values are $13\,\mu$K for $ N=10^4 $, $85 \,\mu$K
for $N=10^6$ and $6200 \,\mu$K for $ N=10^8 $.

\begin{figure}
\centering{\vspace{0.5cm}}
\includegraphics[width=3.5in]{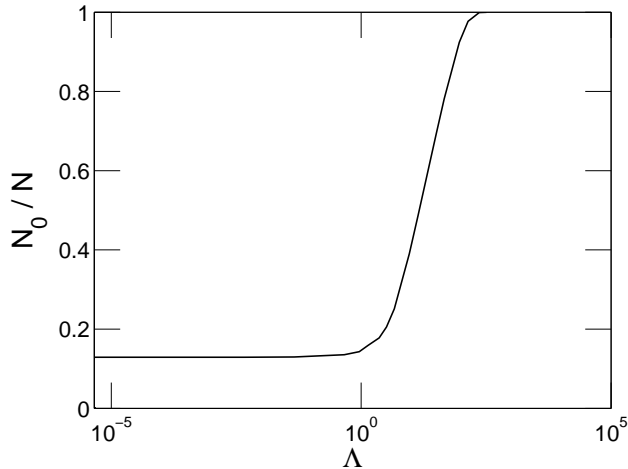}
\caption{Condensate fraction $N_0/N$ vs the strength of the Dirac
\del potential $\Lambda$ when $N=10^4$. The logarithmic scale is
used for the $\Lambda$ axis. The other parameters are the same as
in Fig. \ref{critemp}.}
 \label{condfvss}
\end{figure}

We also find the condensate fraction as a function of  $\Lambda$
at a constant $T=T_c^0$. These results for $N=10^4$ are shown in
Fig. \ref{condfvss}. In the following section we show that the
condensate fraction changes exponentially for large $\Lambda$.
Thus, large $\Lambda$ values ($\Lambda>1$) induce sharp increase
in the condensate fraction.

\begin{figure}
\centering{\vspace{0.5cm}}
\includegraphics[width=3.5in]{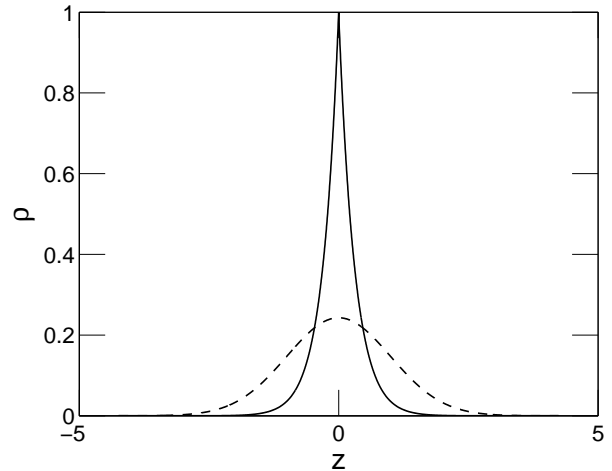}
\caption{Comparison of  density profiles of a BEC in a harmonic
trap with a BEC in a harmonic trap decorated with a \del function
($\Lambda=4.6$). The solid curve is the density profile of the BEC
in decorated potential. The dashed curve is the density profile of
the 1D harmonic trap ($\Lambda=0$). The parameter z is
dimensionless length defined after Eq. (\ref{Scheq}). The other
parameters are the same as in Fig. \ref{critemp}.} \label{Denprof}
\end{figure}

Finally, we compare the density profiles of condensates for a
harmonic trap and a harmonic trap decorated with a \del function
($\Lambda =4.6$) in Fig. \ref{Denprof}. Since the ground-state
wave functions can be calculated analytically for both cases, we
find the density profiles by taking the absolute square of the
ground-state wave functions.

\section{Adiabatic Condensation}

\begin{figure}
\centering{\vspace{0.5cm}}
\includegraphics[width=3.5in]{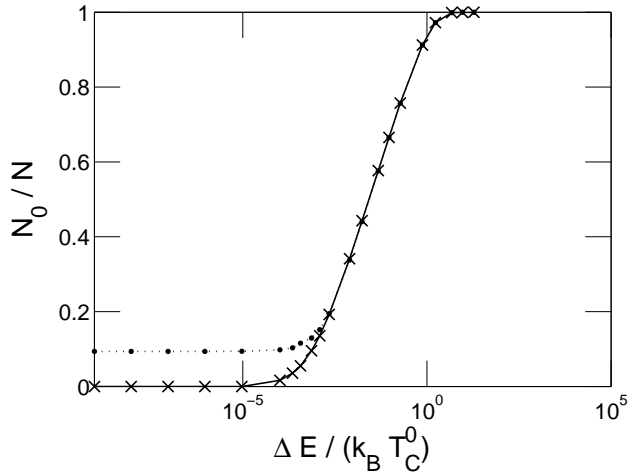}
\caption{The comparison of the condensate fraction values
evaluated with the semiclassical method and our method for
$N=10^4$. The crosses and dots show the results of the
semiclassical method and our method, respectively. The logarithmic
scale is used for the x axis, where $\Delta E=  \hbar \omega / 2
-E_g$.} \label{compCF}
\end{figure}

One can also find the condensate fraction assuming a dimple
potential is added to harmonic trap adiabatically
\cite{kurn,stringari}. We first calculate the grand potential of a
one-dimensional Bose-gas in a harmonic trap using semiclassical
approximation:
\be
\Omega=\Omega_0+\frac{k_B T}{\hbar \omega}  \int_{(3/2) \, \hbar
\omega}^{\infty} d\varepsilon \ln \left[ 1- e^{\beta (\mu-
\varepsilon)} \right]  \label{Grandpotential}
 \ee
where $\Omega_0$ denotes the contribution of the ground state to
the grand potential. The $ 1/(\hbar \omega) $ factor before the
integral in Eq. (\ref{Grandpotential}) is the density of states
for one-dimensional harmonic potential. We calculate the entropy
using $S=-d \Omega/d T$ and get
\bea
 \frac{S}{k_B} & = &   \frac{ 2 k_B T}{\hbar
\omega} g_2 \left( e^{\beta (\mu- \frac{3}{2} \hbar \omega
)}\right)  \nonumber \\ & + & \left( \frac{\mu}{\hbar
\omega}-\frac{3}{2} \right) \ln \left[ 1- e^{\beta (\mu- 3 \hbar
\omega / 2 )} \right] . \label{entropy}
\eea
where $g_2(x)$ is a Bose function  and  is defined as
\be
g_2(x)=\sum_{l=1}^{\infty} \frac{x^l}{l^2} . \label{bosefunctions}
\ee
For temperatures slightly above $T_c^0$, the entropy is
\be
 \frac{S}{k_B}  =  \frac{ 2 k_B T}{\hbar
\omega} g_2 \left( e^{\frac{\hbar \omega}{k_B T_c^0}} \right)- \ln
\left[ 1- e^{-\frac{\hbar \omega}{k_B T_c^0}} \right] ,
\label{entropy2}
\ee
assuming $\mu=\hbar \omega/2$. We assume that adding a dimple
potential changes the ground state and the chemical potential such
that $\mu= E_g < \hbar \omega /2$ and the entropy expression
becomes
\bea
 \frac{S}{k_B} & = & \frac{ 2 k_B T}{\hbar \omega} g_2 \left( e^{\beta (E_g-
\frac{3}{2} \hbar \omega )}\right) \nonumber \\ & + & \left(
\frac{E_g}{\hbar \omega}-\frac{3}{2}  \right) \ln \left[ 1-
e^{\beta (E_g- 3\hbar \omega/2 )} \right] . \label{entropy3}
\eea
Since the entropy remains constant during an adiabatic process, we
can use Eqs. (\ref{entropy2}) and (\ref{entropy3}) to calculate
the final temperature of the system. Then, we calculate the
condensate fraction for this final temperature. We present the
change of condensate fraction  with respect to $\Delta E/(k_B
T_c^0)$ in Fig. \ref{compCF} where $\Delta E= \hbar \omega/2-E_g$.
The solid and dashed lines in Fig. \ref{compCF} show the results
of the semiclassical method and our method, respectively. $\Delta
E$ increases with $\Lambda$, representing the potential depth for
a constant width. We have calculated the condensate fraction for
the same ground-state energy values of the dimple potential in
both the semiclassical and our method to make a comparison. The
condensate fraction results in Fig. \ref{compCF} differ for these
two methods for very small $\Lambda$  values. This is due to two
effects: (a) Calculations of the chemical potential is done
accurately by using the constant N for the {\it finite} number of
particles for our method, however the chemical potential is just
taken the ground-state energy for the semiclassical method
\cite{kurn,stringari}. (b) An accurate summation calculation is
performed for our method; however, the summation is approximated
by an integral for the semiclassical method.

The plots in Figs. \ref{critemp} and \ref{compCF} show that the
agreement of the results of two methods is better for large
$\Lambda$ (or large $\Delta E$) values. This is due to the fact
that the average occupations for excited states is so low ($
n^{excited} \ll 1$) that the semiclassical treatment for this gas
is valid.

\section{Approximate Solutions of The Critical temperature and Condensate fraction for large $\sigma$}

If we apply a deep dimple potential ($\sigma \rightarrow \infty$)
to the atomic condensate in a harmonic trap, the problem can be
solved analytically by approximating the summation for N with an
integral. As $T\rightarrow T_c$, $\mu\rightarrow \varepsilon_0
\approx E_{\delta}$, where
\be E_{\delta}=-\frac{\hbar^2}{2m}(\frac{\sigma}{2})^2 \ee
and it is the bound state energy of a Dirac \del potential
\cite{avakian}. By using density of states $\rho(\varepsilon)$ and
utilizing $E_{2n+2} \rightarrow E_{2n+1}$ (odd-state energies) as
$\sigma \rightarrow \infty$, the summation in Eq. (\ref{chem pot})
is converted to the following integral
\be N=\frac{1}{\hbar \omega}\int_{(3/2)\hbar \omega}^\infty
\frac{d\varepsilon}{e^{\beta_c\varepsilon+\alpha}-1} \ee
where $\alpha=\beta_c (\hbar^2/(2m)) (\sigma/2)^2$ and
$\beta_c=1/k_B T_c$. After calculating the integral, we get
\be
N= -\frac{k_B T_c}{\hbar \omega} \ln\left( 1-\exp\left\{ -
\beta_c\left[ \frac{3 \hbar
\omega}{2}-\frac{\hbar^2}{2m}(\frac{\sigma}{2})^2 \right] \right\}
\right).
\ee
Defining $A=|E_{\delta}|/\hbar\omega$, we have $A \gg 1 $ and
$\exp{((-\beta_c \hbar \omega A))} \ll 1$ for very large $\sigma$.
Then,
\be \label{Number} N\approx\frac{k_B T_c}{\hbar\omega}e^{-\beta_c
\hbar \omega A} \ee
for this case and we get for the critical temperature
\be k_B T_c \approx \frac{A}{\ln(\frac{A/N}{\ln(A/N)})}\hbar\omega
.
 \ee
for large $A/N$ value. For one-dimensional experimental systems,
$N \approx10^3-10^6$ atoms. For a specific case $ A/N \approx 10^4
\approx e^{9}$, one gets
\be k_B T_c \approx \frac{|E_{\delta}|}{7}, \ee
which shows that the critical temperature increases linearly with
increasing bound-state energy for the dimple potential.

By using Eq. (\ref{Number}), the condensate fraction can be
written as
\begin{eqnarray}
\label{newcfrac} \frac{N_0}{N}=1-\frac{T}{T_c} e^{-\beta \beta_c
k_B (T_c-T)\hbar \omega A}
\end{eqnarray}
for $T<T_c$. This result indicates an exponential increase of
$N_0$ as a function of T for $T<T_c$. Thus, the number of atoms in
Bose-Einstein condensate will rise drastically when a very strong,
very short-range (point) interaction is added to a harmonic
confining potential. However, we ignore the interactions between
the atoms and neglect nonlinear terms in the Gross-Pitaevski
equation which will modify these results.
%%%%%%%%%%%%%%%%%%%%%%%%%%%%%%%%%%%%%%%%%%%%%%%%%%%%%%%%%%%%%%%%%%%%
%%%%%%%%%%%%%%%%%%%%%%%%%%%%%%%%%%%%%%%%%%%%%%%%%%%%%%%%%%%%%%%%%%%%
%%%%%%%%%%%%%%%%%%%%%%%%%%%%%%%%%%%%%%%%%%%%%%%%%%%%%%%%%%%%%%%%%%%%

\section{Conclusion}

We have investigated the effect of the tight dimple potential on
the harmonically confined one-dimensional BEC. We model the dimple
potential with the Dirac \del function. This allows for analytical
expressions for the eigenfunctions of the system and a simple
eigenvalue equation greatly simplifying numerical treatment. Pure
analytical results are obtained in the limit of infinitely deep
dimple potential case.

We have calculated the critical temperature, chemical potential
and condensate fraction and demonstrated the effect of the dimple
potential. We have found that the critical temperature can be
enhanced by an order of magnitude for experimentally accessible
dimple potential parameters. In general, $T_c$ increases with the
relative strength of the dimple potential with respect to the
harmonic trap. In our model system, the increase in the strength
of the Dirac \del function can be interpreted as increasing the
depth of a dimple potential.

In addition, the change of the condensate fraction with respect to
the strength of the Dirac \del function has been analyzed at a
constant temperature ($T=T_c^0$) and with respect to temperature
at a constant strength. It has been shown that the condensate
fraction can be increased considerably and large condensates can
be achieved at higher temperatures due to the strong localization
effect of the dimple potential. Analytical expressions are given
to clarify the relation of the condensate fraction and critical
temperature to the strength of deep dimple potential.

We also show that similar results for the critical temperature and
condensate fraction can be obtained by using a semiclassical
approximation. However, in this simpler approach, it is not
possible to determine the density profile and when the
thermodynamic limit is not satisfied, the estimations of
semiclassical approach fail.

Finally, we have determined and compared the density profiles of
the harmonic trap and the decorated trap with the Dirac \del
function at the equilibrium point using analytical solutions of
the model system. Comparing the graphics of density profiles, we
see that a dimple potential maintains a considerably higher
density at the center of the harmonic trap.

The results presented are obtained for the case of noninteracting
condensates for simplicity. The treatment should be extended to
the case of interacting condensates in order to make the results
more relevant to experimental investigations. This case deserves
further detailed and separate calculations. We believe the method
presented of $\delta$-function modelling of tight dimple
potentials can help significantly such theoretical examinations.

%%%%%%%%%%%%%%%%%%%%%%%%%%%%%%%%%%%%%%%%%%%%%%%%%%%%%%%%%%%%%%%%%%%%
%%%%%%%%%%%%%%%%%%%%%%%%%%%%%%%%%%%%%%%%%%%%%%%%%%%%%%%%%%%%%%%%%%%%
%%%%%%%%%%%%%%%%%%%%%%%%%%%%%%%%%%%%%%%%%%%%%%%%%%%%%%%%%%%%%%%%%%%%

\acknowledgments

O.E.M. acknowledges support from a T\"UBA/GEB\.{I}P grant. E.D. is
supported by Turkish Academy of Sciences, in the framework of the
Young Scientist Program (ED- T\"UBA- GEBIP-2001-1-4).

%%%%%%%%%%%%%%%%%%%%%%%%%%%%%%%%%%%


\begin{thebibliography}{99}
\bibitem{mistake} There is a small numerical error in \cite{avakian}. The right
hand side of Eq. (12) in \cite{avakian} should be divided by
$2^{3/2}$.
\end{thebibliography}
\end{document}